\documentstyle[12pt,epsfig]{article}

\catcode`@=11
\def\marginnote#1{}
\newcount\hour
\newcount\minute
\newtoks\amorpm
\hour=\time\divide\hour by60
\minute=\time{\multiply\hour by60 \global\advance\minute by-\hour}
\edef\standardtime{{\ifnum\hour<12 \global\amorpm={am}%
        \else\global\amorpm={pm}\advance\hour by-12 \fi
        \ifnum\hour=0 \hour=12 \fi
        \number\hour:\ifnum\minute<10 0\fi\number\minute\the\amorpm}}
\edef\militarytime{\number\hour:\ifnum\minute<10 0\fi\number\minute}
\def\draftlabel#1{{\@bsphack\if@filesw {\let\thepage\relax
   \xdef\@gtempa{\write\@auxout{\string
      \newlabel{#1}{{\@currentlabel}{\thepage}}}}}\@gtempa
   \if@nobreak \ifvmode\nobreak\fi\fi\fi\@esphack}
        \gdef\@eqnlabel{#1}}
\def\@eqnlabel{}
\def\@vacuum{}
\def\draftmarginnote#1{\marginpar{\raggedright\scriptsize\tt#1}}
\def\draft{\oddsidemargin 0.0truein
        \def\@oddfoot{\sl preliminary draft \hfil
        \rm\thepage\hfil\sl\today\quad\militarytime}
        \let\@evenfoot\@oddfoot \overfullrule 3pt
        \let\marginnote=\draftmarginnote
   \def\@eqnnum{(\theequation)\rlap{\kern\marginparsep\tt\@eqnlabel}%
\global\let\@eqnlabel\@vacuum}  }
\def\slash#1{\setbox0=\hbox{$#1$}#1\hskip-\wd0\dimen0=5pt\advance
       \dimen0 by-\ht0\advance\dimen0 by\dp0\lower0.5\dimen0\hbox
         to\wd0{\hss\sl/\/\hss}}
\catcode`@=12

\def\thefootnote{\arabic{footnote}}
\makeatletter

\def\secteqno{\@addtoreset{equation}{section}%
\def\theequation{\thesection.\arabic{equation}}}

\def\endsecteqno{\def{theequation\{\@ifundefined{chapter}%
{\arabic{equation}}{\thechapter.\arabic{equation}}}}

\makeatletter

\newcommand{\s}{\\ \vspace {-3mm} }

\newcommand{\lsim}{\raisebox{-0.13cm}{~\shortstack{$<$ \\[-0.07cm] $\sim$}}~}

\newcommand{\beq}{\begin{eqnarray}}
\newcommand{\eeq}{\end{eqnarray}}
\newcommand{\ba}{\begin{array}}
\newcommand{\ea}{\end{array}}

\def\ijmp#1#2#3{{\it Int. Jour. Mod. Phys. }{\bf #1~}(19#2)~#3}

\def\plb#1#2#3{{\it Phys. Lett. }{\bf B#1~}(19#2)~#3}
\def\zpc#1#2#3{{\it Z. Phys. }{\bf C#1~}(19#2)~#3}
\def\prl#1#2#3{{\it Phys. Rev. Lett. }{\bf #1~}(19#2)~#3}
\def\rmp#1#2#3{{\it Rev. Mod. Phys. }{\bf #1~}(19#2)~#3}
\def\prep#1#2#3{{\it Phys. Rep. }{\bf #1~}(19#2)~#3}
\def\prd#1#2#3{{\it Phys. Rev. }{\bf D#1~}(19#2)~#3}
\def\npb#1#2#3{{\it Nucl. Phys. }{\bf B#1~}(19#2)~#3}

\def\arnps#1#2#3{{\it Annu. Rev. Nucl. Part. Sci. }{\bf #1~}(19#2)~#3}
\def\sjnp#1#2#3{{\it Sov. J. Nucl. Phys. }{\bf #1~}(19#2)~#3}

\def\ptp#1#2#3{{\it Prog. Theor. Phys. }{\bf #1~}(19#2)~#3}
\def\hepph#1{{\bf hep-ph}/#1}

\begin{document}
\setlength{\unitlength}{1cm}
\sloppy
%\secteqno

%%%%%%%%%%%%%%%%%%%%%%%%%%%%%%%%%%%%%%%%%%%%%%%%%%%%%%%%%%%%%%%%%%%%%%
% FINAL VERSION							     %
%%%%%%%%%%%%%%%%%%%%%%%%%%%%%%%%%%%%%%%%%%%%%%%%%%%%%%%%%%%%%%%%%%%%%%
% \draft
%%%%%%%%%%%%%%%%%%%%%%%%%%%%%%%%%%%%%%%%%%%%%%%%%%%%%%%%%%%%%%%%%%%%%%

\begin{titlepage}

\begin{flushright}
KA--TP--21--1997\\
DFPD/97/45 \\
{\tt hep-ph/9711322}\\
November 1997 
\end{flushright}

\def\thefootnote{\fnsymbol{footnote}}

\vspace{1cm}

\begin{center}

{\large\sc {\bf Weak Electric Dipole Moments of Heavy Fermions in the MSSM}}

\vspace{1cm}

{\sc W.~Hollik${}^{a}$,
J.I.~Illana${}^a$,
S.~Rigolin${}^{a,b}$,
D.~St\"ockinger${}^{a}$} 
\footnote{E--mail addresses: 
\{hollik,jillana,ds\}@itpaxp3.physik.uni-karlsruhe.de, rigolin@pd.infn.it}  

\vspace{1cm}
${}^a$
Institut f\"ur Theoretische Physik, Universit\"at Karlsruhe,\\
D--76128 Karlsruhe, FR Germany 
\\\vspace{.3cm}
${}^b$
Dipartimento di Fisica, Universit\`a di Padova and INFN,\\
I--35131 Padua, Italy \\

\end{center}

\vspace{1cm}

\begin{abstract}

A minimal supersymmetric version of the Standard Model with complex parameters
allows contributions to the weak--electric dipole moments of fermions
at the one--loop level. 
Assuming generation--diagonal trilinear soft--susy--breaking
terms and the usual GUT constraint, a set of CP--violating physical phases
can be introduced. In this paper the general expressions for the one--loop
contribution to the WEDM in a generic renormalizable theory are given and
the size of the WEDM of the $\tau$ lepton and the $b$ quark in such a 
supersymmetric model is discussed.

\end{abstract}

\end{titlepage}

\setcounter{page}{2}
\setcounter{footnote}{0}
%
%%%%%%%%%%%%%%%%%%%%%%%%%%%%%%%%%%%%%%%%%%%%%%%%%%%%%%%%%%%%%%%%%%%%%%%%%%%
%
%\section*{Introduction}

In the electroweak Standard Model (SM) there is only one source of CP 
violation, the $\delta_{\rm CKM}$ phase of the 
Cabibbo--Kobayashi--Maskawa mixing matrix for quarks \cite{ckm}.\footnote{
The QCD Lagrangian includes an additional source of CP violation, the
$\theta_{\rm QCD}$, but we will not consider it here. Extreme fine tuning
is needed in order that its contribution to the neutron EDM does not exceed 
the present experimental upper bound. Various mechanisms beyond the SM have
been proposed to solve this problem \cite{cpreview}.} 
The only place where CP violation has been currently measured, the neutral 
$K$ system, fixes the value of this phase but does not constitute itself a 
test for the origin of CP violation \cite{cpreview}.
On the other hand, if the baryon asymmetry of the universe has been dynamically
generated, CP must be violated. The SM cannot account for the 
size of the observed asymmetry \cite{b-asym}. Many extensions of the SM contain
new CP--violating phases, in particular, the supersymmetric models 
\cite{susycp}. 
It has also been shown that the Minimal Supersymmetric Standard Model (MSSM)
\cite{mssm} can provide the correct size of baryon asymmetry in some
range of parameters if the CP--violating phases are not suppressed 
\cite{b-susyasym}.\s

One needs soft--breaking terms to introduce physical phases in the MSSM, 
different from the $\delta_{\rm CKM}$ \cite{abelfrere}. 
We assume that the soft--breaking terms preserve R--parity.
Other possibilities for CP violation can arise in R--parity violating models
(cf. e.g. \cite{abel} in the context of R--parity violating scalar 
interactions). For simplicity, we restrict ourselves to 
generation--diagonal trilinear soft--breaking terms to prevent FCNC.
Doing this we ignore CP--violating effects that have been already considered 
in the literature \cite{fcnc-cp}.
In our analysis we do not make any additional assumption, except
for the unification of the soft--breaking gaugino masses at the GUT scale. 
However we do not assume unification of the scalar mass parameters or trilinear 
mass parameters.
In such a constrained framework the following SUSY parameters can be complex:
the Higgs--Higgsino mass parameter $\mu$; the gaugino mass parameters 
$M_1$, $M_2$ and
$M_3$; the bilinear mixing mass parameter $m^2_{12}$ and the trilinear
soft supersymmetry breaking parameters $A_\tau$, $A_t$ and $A_b$ (and
accordingly for the other two generations). Not all of these phases are
physical. Namely, the MSSM has two additional U(1) symmetries for vanishing 
$\mu$ and soft--breaking terms: the Peccei--Quinn and the R--symmetry. For
non vanishing $\mu$ and soft--breaking terms these symmetries can
be used to absorb two of the phases by redefinition of the fields 
\cite{relax}. In addition, the GUT relationship between $M_1$, $M_2$ and
$M_3$
leads to only one common phase
for the gaugino mass parameters. Hence, we are left with
four independent CP--violating physical phases (only two assuming 
universality in the sfermion sector). \s

The most significant effect of the CP--violating phases in the phenomenology is 
their contribution to electric dipole moments (EDMs) \cite{dipoles}. Unlike the
SM, where the contribution to the EDM of fermions arises beyond two loops 
\cite{smedm}, the MSSM can give a contribution already at the one--loop level.
The measurement of the neutron EDM \cite{edm-n} constrains the phases and
the supersymmetric spectrum in a way that may demand fine tuning (supersymmetric
CP problem): phases of ${\cal O}(10^{-2})$ \cite{susycp}, or SUSY 
particles very heavy (several TeV \cite{nath1}). 
This problem could be solved if 
soft supersymmetry breaking terms are universal and the genuine SUSY CP phases 
vanish (the Yukawa matrices are then the only source of CP violation, like in 
the SM). It has been argued that one could still keep the SUSY phases of 
${\cal O}(1)$ and the SUSY spectrum not very heavy and satisfy the experimental
bounds due to cancellations among the different components of the neutron EDM
\cite{nath2}. Furthermore it has been recently shown \cite{atwood} that, even 
without such cancellations and in the context of non universal soft--breaking 
terms, the current experimental limits on the neutron EDM can
be met with almost no fine tunning on the CP--violating phases (even for
the first generation ones), at the only price of arg$(\mu)$ of ${\cal O}
(10^{-2})$.
In our analysis we do not assume universality and keep all the SUSY phases as 
free parameters.\s

As a generalization of the electromagnetic dipole moments of fermions, one can 
define weak dipole moments (WDMs), corresponding to couplings with a $Z$ boson
instead of a photon.
The most general Lorentz structure of the vertex function that 
couples a $Z$ boson and two on--shell fermions (with outgoing momenta
$q$ and $\bar{q}$) can be written in terms of form factors
$F_i(s\equiv(q+\bar{q})^2)$ as 
\beq
\Gamma^{Zff}_\mu&=&ie\Bigg\{\gamma_\mu\left[\left(
F_{\rm V}-\frac{v_f}{2s_Wc_W}\right)-\left(F_{\rm A}-\frac{a_f}{2s_Wc_W}\right)
\gamma_5\right] \nonumber \\ & & +(q-\bar{q})_\mu[F_{\rm M}+F_{\rm E}\gamma_5] 
-(q+\bar{q})_\mu[F_{\rm S}+F_{\rm P}\gamma_5]\Bigg\}\ ,
\label{vertex}
\eeq
where $v_f\equiv(I^f_3-2s^2_WQ_f)$, $a_f\equiv I^f_3$.
The form factors $F_{\rm M}$ and $F_{\rm E}$ are related to the anomalous
weak magnetic and electric dipole moments of the fermion $f$ with mass $m_f$ 
as follows:
\beq
{\rm AWMDM}\equiv a^W_f&=&-2m_f\ F_{\rm M}(M^2_Z) \ , \nonumber\\
{\rm WEDM}\equiv d^W_f&=&ie\ F_{\rm E}(M^2_Z) \ .\nonumber
\eeq
The $F_{\rm M}$ ($F_{\rm E}$) form factors are the coefficients 
of the {\em chirality--flipping} term of the CP--conserving (CP--violating) 
effective Lagrangian describing $Z$--fermion couplings. Therefore, they 
are expected to get contributions proportional to some positive power of 
the mass of the fermions involved. This allows the construction of observables
which can be probed experimentally most suitably by heavy fermions. Hence,
for on--shell $Z$ bosons, where the dipole form factors are gauge independent,
the $b$ quark and $\tau$ lepton are the most promising candidates.\s

In this work we concentrate on the analysis of the one--loop
contribution of the MSSM with complex parameters and conserved R--parity 
to the WEDM of the $\tau$ lepton and the $b$ quark.\footnote{
The AWMDM has been considered in Refs.~\cite{hirs} where real supersymmetric
couplings were used.}

\section*{The WEDM}

\begin{figure}[htb]
\begin{center}
\begin{picture}(15,6)
\epsfxsize=14cm
\put(0.5,0){\epsfbox{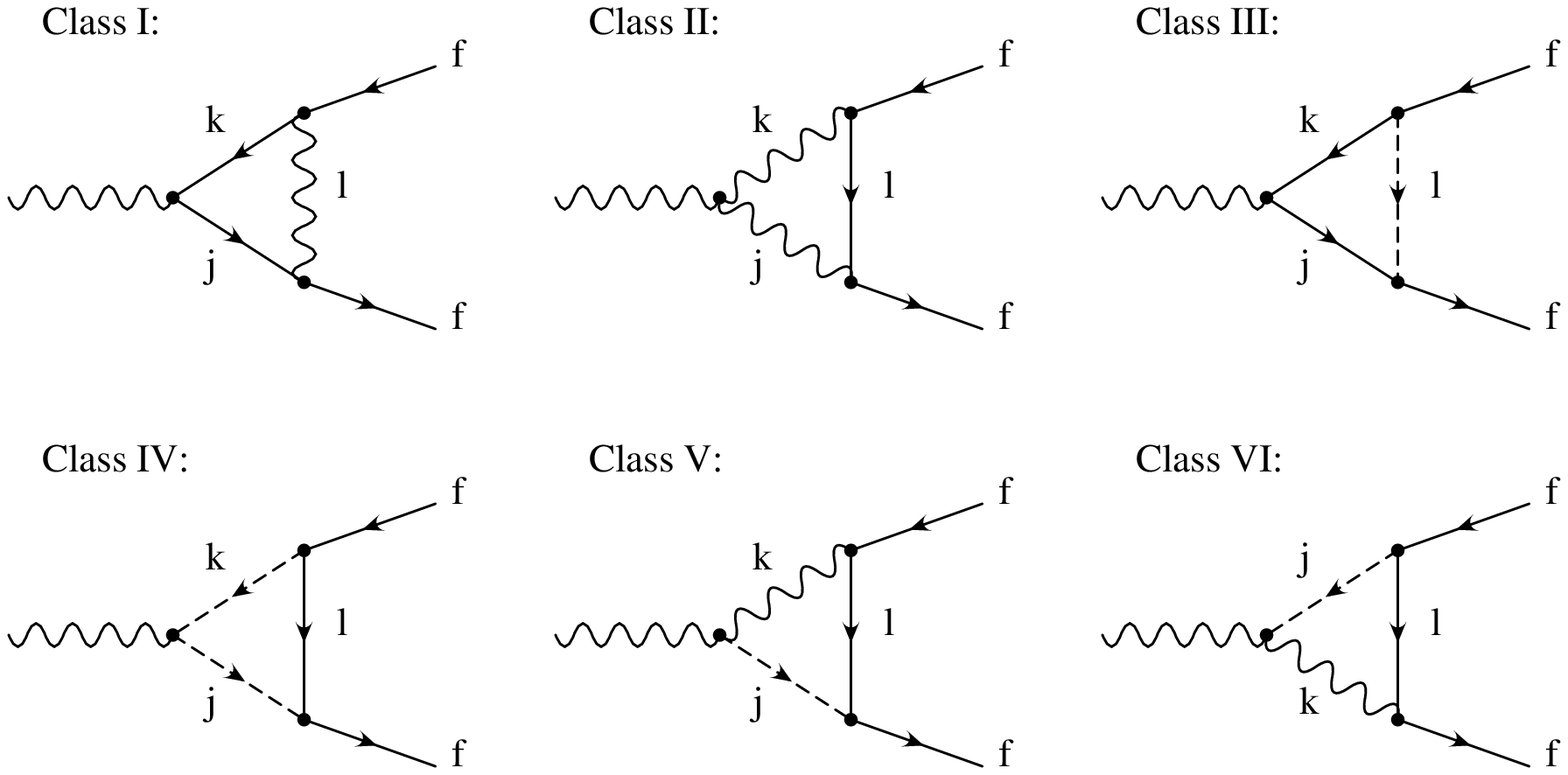}}
\end{picture}
\end{center}
\caption{\sl \small The one--loop $Zff$ diagrams.
\label{fig1}}
\end{figure}

All the possible one--loop contributions to the WEDM can be classified in
terms of six classes of triangle diagrams (Fig.~\ref{fig1}). 
The vertices are labelled by generic couplings, according to the following 
interaction Lagrangian, for vectors $V^{(k)}_\mu=A_\mu,\ Z_\mu,\ W_\mu,
\ W^\dagger_\mu$, general fermions $\Psi_k$ and general scalars $\Phi_k$:
\beq
{\cal L}&=&ieJ (W^\dagger_{\mu\nu}W^\mu Z^\nu-W^{\mu\nu}W^\dagger_\mu Z_\nu
             +Z^{\mu\nu}W^\dagger_\mu W_\nu)
 + e V^{(k)}_\mu\bar{\Psi}_j\gamma^\mu(V^{(k)}_{jl}-A^{(k)}_{jl}\gamma_5)\Psi_l
\nonumber \\
 &+&ieG_{jk} Z^\mu\Phi_j^\dagger\stackrel{\leftrightarrow}{\partial}_\mu\Phi_k
     + \Big\{ e \bar{\Psi}_f(S_{jk}-P_{jk}\gamma_5)\Psi_k\Phi_j\ 
     + eK_{jk} Z^\mu V^{(k)}_\mu\Phi_j + h.c.\Big\}
\eeq
The expressions for the WEDM are evaluated in the 't Hooft--Feynman gauge
(all the would--be--Goldstone bosons must be included) and are written in
terms of three--point one--loop tensor integrals $\bar{C}$ and vertex
coefficients:\footnote{
Equivalent expressions for classes III and IV can be found in 
Ref.~\cite{vienna} where a different set of generic couplings and tensor 
integrals is employed.}
\beq
\frac{d^W_f}{e}({\rm I})=\frac{\alpha}{4\pi}  \Big\{ 
4m_f\sum_{jkl}{\rm Im}
             [V^Z_{jk}(V^{(l)}_{fj}A^{(l)*}_{fk}+A^{(l)}_{fj}V^{(l)*}_{fk})
             +A^Z_{jk}(V^{(l)}_{fj}V^{(l)*}_{fk}+A^{(l)}_{fj}A^{(l)*}_{fk})] 
        \nonumber\\
\times [2C^{+-}_2-C^-_1](k,j,l) \hspace{2cm}    \nonumber\\
-4\sum_{jkl}m_k{\rm Im}
             [V^Z_{jk}(V^{(l)}_{fj}A^{(l)*}_{fk}-A^{(l)}_{fj}V^{(l)*}_{fk})
             -A^Z_{jk}(V^{(l)}_{fj}V^{(l)*}_{fk}-A^{(l)}_{fj}A^{(l)*}_{fk})] 
         \nonumber\\
\times [2C^{+}_1-C_0](k,j,l) \Big\}
\hspace{2cm} 
\eeq
\beq
\frac{d^W_f}{e}({\rm II})=-\frac{\alpha}{4\pi} \Big\{ 
2m_f\sum_{jkl}{\rm Im}[J(V^{(j)}_{fl}A^{(k)*}_{fl}+A^{(j)}_{fl}V^{(k)*}_{fl})]
[4C^{+-}_2-C^-_1](k,j,l)
\nonumber\\
+6\sum_{jkl} m_l{\rm Im}[J(V^{(j)}_{fl}A^{(k)*}_{fl}-A^{(j)}_{fl}V^{(k)*}_{fl})]
C^+_1(k,j,l) 
\Big\}\hspace{1cm}
\eeq
\beq
\frac{d^W_f}{e}({\rm III})=\frac{\alpha}{4\pi}  \Big\{
-2m_f\sum_{jkl}
{\rm Im}[V^Z_{jk}(P_{lj}S^*_{lk}+S_{lj}P^*_{lk})
        +A^Z_{jk}(S_{lj}S^*_{lk}+P_{lj}P^*_{lk})] \nonumber\\
\times [2C^{+-}_2-C^-_1](k,j,l) \hspace{2cm}    \nonumber\\
+2\sum_{jkl} {m}_k 
{\rm Im}[V^Z_{jk}(P_{lj}S^*_{lk}-S_{lj}P^*_{lk})
        +A^Z_{jk}(S_{lj}S^*_{lk}-P_{lj}P^*_{lk})] \nonumber\\
\times [C^+_1+C^-_1](k,j,l)       \Big\}
\hspace{2cm}
\label{III}
\eeq
\beq
\frac{d^W_f}{e}({\rm IV})=\frac{\alpha}{4\pi}  \Big\{
2m_f\sum_{jkl}
{\rm Im}[G_{jk}(S_{jl}P^*_{kl}+P_{jl}S^*_{kl})][2C^{+-}_2-C^-_1](k,j,l)
\nonumber\\
-2\sum_{jkl} {m}_l 
{\rm Im}(G_{jk}S_{jl}P^*_{kl})[2C^+_1-C_0](k,j,l)       \Big\}
\hspace{1.6cm}
\label{IV}
\eeq
\beq
\frac{d^W_f}{e}({\rm V+VI})=-\frac{\alpha}{4\pi} \sum_{jkl} 2 
{\rm Im}[K_{jk}(V^{(k)}_{fl}P_{jl}^*+A^{(k)}_{fl}S_{jl}^*)][C^+_1+C^-_1](k,j,l) 
\eeq
The arguments of the tensor integrals refer to $\bar{C}(k,j,l)\equiv 
\bar{C}(-\bar{q},q,M_k,M_j,M_l)$ of Ref.~\cite{topol}.
The tensor integrals are defined in such a way that for equal external fermion
masses $C^-_1$ and $C^{+-}_2$ are antisymmetric under the interchange of $k$ and
$j$, whereas $C_0$ and $C^+_1$ are symmetric. The contribution of diagrams of 
class I and II vanishes as they can only involve SM fermions in the loop 
(MSSM preserves R--parity) whose couplings to gauge
bosons are either real ($Z$--exchange) or self--conjugated ($W$--exchange). 
The gluonic contribution in class I contains only real couplings. 
For the class V and VI diagrams, the only contribution to the WEDM might
occur when a pseudoscalar Higgs boson is involved in the loop, but 
there is no coupling of two neutral gauge bosons to a pseudoscalar
and hence they also vanish. One can easily check that the Higgs sector 
for both the SM and the MSSM to class III and IV diagrams does not contribute
to the WEDM, consistently with the CP--conserving character of both the SM and
MSSM Higgs sectors. In a general 2HDM a CP--violating contribution is possible 
\cite{2hdm-cp}. These considerations lead to the well known result that
the SM one--loop contribution to the WEDM is zero. The MSSM contribution 
comes from charginos, neutralinos, gluinos and sfermions via diagrams of 
class III and IV. Finally, notice that all the contributions are proportional
to one of the fermion masses involved, as expected from the 
chirality flipping character
of the dipole moments.\footnote{For the diagrams of class V and VI the
chirality flipping occurs at the scalar--fermion vertex and the
fermion mass is embedded in $S$ and $P$, which are in this 
case Yukawa couplings.}

\section*{The WEDM of \boldmath$\tau$ lepton and the \boldmath$b$ quark}

The conventions for couplings and mixings in the MSSM are the ones in 
Ref.~\cite{mssm,mssmdetailed} except for the complex character of the
$\mu$ parameter and the trilinear soft supersymmetry breaking parameters 
$A_\tau$, $A_t$ and $A_b$. For convenience, we deal with the following 
CP--violating phases:\footnote{
Such a choice leads to a dependence on $\varphi_\mu$ of chargino and 
neutralinos masses. Conversely the sfermion masses are independent on
$\varphi_{\tilde{f}}$.}
$\varphi_\mu\equiv{\rm arg}(\mu)$, 
$\varphi_{\tilde{f}}\equiv{\rm arg}(m^f_{\rm LR})$ ($f=\tau,\ t,\ b$)
with $m^t_{\rm LR}\equiv A_t-\mu^*\cot\beta$ and 
$m^{\tau,b}_{\rm LR}\equiv A_{\tau,b}-\mu^*\tan\beta$. 
We assume a common squark mass parameter $m_{\tilde{q}}\equiv m_{\tilde{Q}}=
m_{\tilde{U}}=m_{\tilde{D}}$ as well as a common slepton mass parameter
$m_{\tilde{l}}\equiv m_{\tilde{L}}=m_{\tilde{E}}$. We take real gaugino mass 
parameters constrained by the GUT relations: 
\beq
M_1=\frac{5}{3}\tan^2\theta_W M_2\ , \ \ \ \ 
M_3=\frac{\alpha_s}{\alpha} s^2_W M_2\ .
\label{gut}
\eeq

A ``natural" scale for the EDMs is a ``magneton'' defined by $\mu_f\equiv 
e/2m_f=1.7\times10^{-15}\ (0.7\times10^{-15})\ e$cm for the $\tau$ lepton 
($b$ quark). In the plots the dimensionless quantity $d^W_f/\mu_f$ is 
displayed.\s

We make a full scan of the SUSY parameter space and determine the values
of the CP--violating phases that yield the maximum effect on the WEDM. 
The result of this analysis is described below.
 
\subsubsection*{Chargino and scalar neutrino contribution to 
\boldmath$d^W_\tau$}

There is only one phase, $\varphi_\mu$, involved in the chargino contribution
as there is no mixing in the scalar neutrino sector. The result grows with
$\tan\beta$. It also depends on the common slepton mass (whose effect 
consists of dumping the result through the tensor integrals) and the $|\mu|$ 
and $M_2$ mass parameters. A value $\varphi_\mu=\pi/2$ enhances the WEDM.
Taking $M_2=|\mu|=250$ GeV, Re$(d^W_\tau$[charginos]$)=0.18\ (5.52)\times 
10^{-6}\ \mu_\tau$ for $\tan\beta=1.6\ (50)$ and $m_{\tilde{l}}=250$ GeV.  
There is no contribution to the imaginary part assuming the present bounds on 
the chargino masses.

\subsubsection*{Neutralino and \boldmath$\tilde\tau$ slepton contribution to 
\boldmath$d^W_\tau$}

Now both $\varphi_\mu$ and $\varphi_{\tilde{\tau}}$ contribute. Assuming
that $|m^\tau_{LR}|$ is of the order of $|\mu|\tan\beta$ or below we get 
that there is no large influence of $\varphi_{\tilde{\tau}}$ on the neutralino
contribution,\footnote{The size of $|m^\tau_{LR}|$ is critical for the $\tau$ 
sleptons to have a physical mass.} for both low and high $\tan\beta$. Thus
we observe that this contribution is roughly proportional to $\sin\varphi_\mu$
regardless the value of $\varphi_{\tilde{\tau}}$. Taking $\varphi_\mu=\pi/2$,
$M_2=|\mu|=250$ GeV and $m_{\tilde{l}}=250$ GeV we find that the 
Re$(d^W_\tau$[neutralinos]$)=-0.01\ (-0.25)\times 10^{-6}\ \mu_\tau$ for 
$\tan\beta=1.6\ (50)$.
For presently non excluded masses of the neutralinos there can be a contribution
to the imaginary part, of the order of $10^{-6}\ \mu_\tau$.

\subsubsection*{Chargino and \boldmath$\tilde t$ squark contribution to 
\boldmath$d^W_b$}

Two CP--violating phases are involved in this contribution:
$\varphi_\mu$ and $\varphi_{\tilde{t}}$. In Fig.~2(a) the dependence on these
phases is shown, for $M_2=|\mu|=m_{\tilde{q}}=250$ GeV, $|m^t_{LR}|=|\mu|
\cot\beta$ and both low and high $\tan\beta$ scenarios. The maximum effect 
on the WEDM is obtained for $\varphi_\mu=\pi/2$ and $\varphi_{\tilde{t}}=\pi$.
For example, one gets Re$(d^W_b$[charginos]$)=
1.17\ (27.1)\times 10^{-6}\ \mu_b$ for low (high) $\tan\beta$.
As expected, in the high $\tan\beta$ scenario our assumed $|m^t_{LR}|$ 
takes a small value and the dependence on $\varphi_{\tilde{t}}$ tends to
dissappear. To have an idea of the maximum value achievable for the
chargino contribution, we show in Fig.~3 the dependence on $M_2$ and $|\mu|$
for $\varphi_\mu=\pm\pi/2$ and $m^t_{LR}=0$ with the previous value for the
common squark mass parameter.  

\subsubsection*{Neutralino and \boldmath$\tilde b$ squark contribution to 
\boldmath$d^W_b$}

The two relevant CP--violating phases for this case are: 
$\varphi_\mu$ and $\varphi_{\tilde{b}}$. As before, the most important
effect from $\varphi_{\tilde{b}}$ arises when the off--diagonal term is
larger, which in this case corresponds to high $\tan\beta$, as the trilinear 
soft breaking parameter is taken to be of the order of $|\mu|\tan\beta$.
The maximum value for the neutralino contribution occurs for 
$\varphi_\mu=\varphi_{\tilde{b}}=\pi/2$ (Fig.~2(b)). The total contribution 
increases with $\tan\beta$. Thus one gets Re$(d^W_b$[neutralinos]$)=-0.29\
(-12.6) \times 10^{-6}\ \mu_b$ for low (high) $\tan\beta$ with
$M_2=|\mu|=m_{\tilde{q}}=250$ GeV and $|m^b_{LR}|=|\mu|\tan\beta$.

\subsubsection*{Gluino contribution to \boldmath$d^W_b$}

The gluino contribution is affected only by $m^b_{LR}$, $m_{\tilde{q}}$ and 
the gaugino mass $M_3$. Therefore
the maximum value occurs for $\varphi_{\tilde{b}}=\pi/2$. 
The mixing in the $\tilde b$ sector is determined by $m^b_{LR}$ and intervenes
in the contribution due to chirality flipping in
the gluino internal line (the contribution proportional to $M_3$). 
The contribution to 
the AWMDM is enhanced by the largest values of 
$|m^b_{LR}|$ compatible with an experimentally not excluded mass 
for the lightest $\tilde b$ squark. 
For zero gluino mass, only the term proportional to the mass
of the $b$ quark provides a contribution. As we increase the gluino mass, 
the term proportional to $M_3$ dominates, especially for large 
$|m^b_{LR}|$, being again suppressed at high $M_3$ due to the gluino 
decoupling. Thus for $\varphi_{\tilde{b}}=\pi/2$ one gets
Re$(d^W_b$[gluinos]$)=0.26\ (9.31)\times 10^{-6}\ \mu_b$ for 
low (high) $\tan\beta$ and $|m^b_{LR}|=|\mu|\tan\beta$, 
$M_2=|\mu|=m_{\tilde{q}}=250$ GeV and $M_3$ fulfilling the GUT relation 
(\ref{gut}).

\section*{Conclusions}

Unlike the SM, an R--parity preserving MSSM version with complex parameters 
contains enough freedom 
to provide a contribution to the (W)EDMs to one loop:
considering generation--diagonal trilinear soft--susy--breaking terms, to reduce 
undesired FCNC, and
the GUT constraint between the gauginos mass parameters, at most two (three) 
CP--violating physical phases are available for lepton (quark) WEDMs, 
apart from the $\delta_{\rm CKM}$.\s

In this work, 
the one--loop analytical expressions for the (W)EDM of fermions in any 
renormalizable theory are given in terms of a set of generic couplings.
Moreover, a full scan of the 
MSSM parameter space has been performed in search for the maximum effect
on the WEDM of the $\tau$ lepton and the $b$ quark. The Higgs sector does
not contribute and chargino diagrams are more important than neutralino
ones. Gluinos are also involved in the $b$ case and compete in importance
with charginos. In the most favourable
configuration of CP--violating phases and for values of the rest of the
parameters still not excluded by experiments, these WEDMs can be as much
as twelve orders of magnitude larger than the SM predictions, although
still far from experimental reach:
\beq
|{\rm Re}(d^W_\tau)|&\lsim&0.3\ (10)\times10^{-21}\ e{\rm cm} \nonumber\\
|{\rm Re}(d^W_b)|&\lsim&\ \ 1\ (20)\times10^{-21}\ e{\rm cm} \nonumber
\eeq
There may be a contribution to the imaginary part if the neutralinos are light.
The current experimental bound on the $\tau$ WEDM \cite{opal} is 
$|$Re$(d^W_\tau)|<5.6\times10^{-18}\ e$cm and $|$Im$(d^W_\tau)|
<1.5\times10^{-17}\ e$cm at 95\% confidence level. There does not exist
any similar analysis for the $b$ which might be not possible due to
hadronization.
For comparison with other theoretical predictions: the electron EDM in the SM 
\cite{dipoles} ($d_f\propto m_f$) can be estimated to be $d_e\sim10^{-37}\ e$cm;
in multi--Higgs models \cite{multihiggs} 
$d^W_\tau\sim 3\times10^{-22}\ e$cm;
in leptoquark models \cite{leptoquarks}
$d^W_\tau\sim 10^{-19}\ e$cm.

\section*{Acknowledgements}

We thank T. Gajdosik for discussions on Ref.~\cite{vienna} and T. Hahn for
valuable help in the preparation of the figures.
J.I.I. is supported by the Fundaci\'on Ram\'on
Areces and partially by the spanish CICYT under contract AEN96-1672. S.R. is 
supported by the Fondazione Ing. A. Gini and by the italian MURST.

%
%%%%%%%%%%%%%%%%%%%%%%%%%%%%%%%%%%%%%%%%%%%%%%%%%%%%%%%%%%%%%%%%%%%%%%%%%%%
%

%
%%%%%%%%%%%%%%%%%%%%%%%%%%%%%%%%%%%%%%%%%%%%%%%%%%%%%%%%%%%%%%%%%%%%%%%%%%%
%
\newpage
\section*{Figures}
\vspace{1cm}

\begin{figure}[htb]
\begin{center}
\begin{picture}(15,6.75)
\epsfxsize=8cm
\put(-0.8,-0.2){\epsfbox{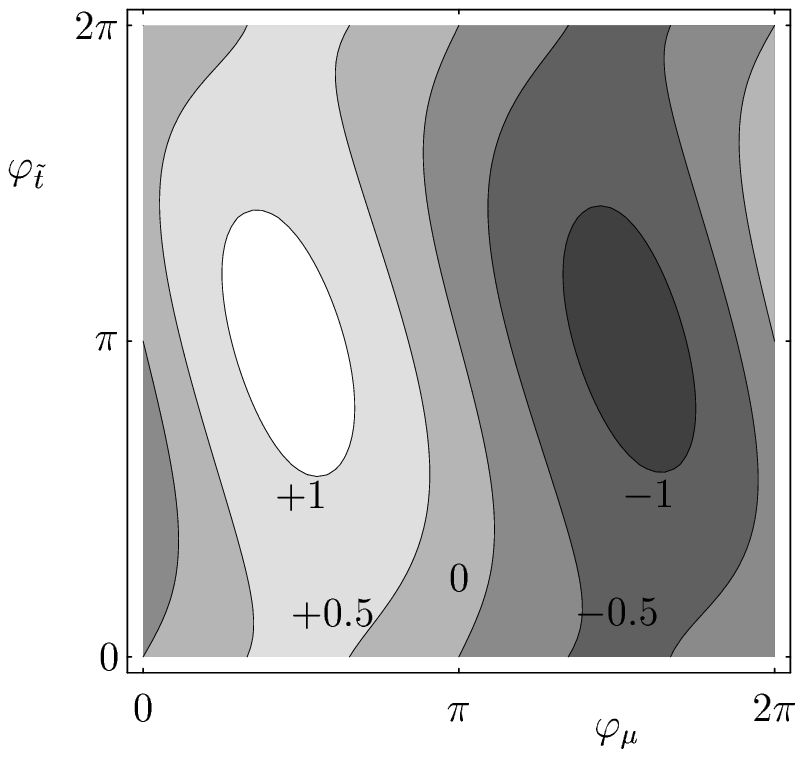}}
\epsfxsize=8cm
\put(7.7,-0.2){\epsfbox{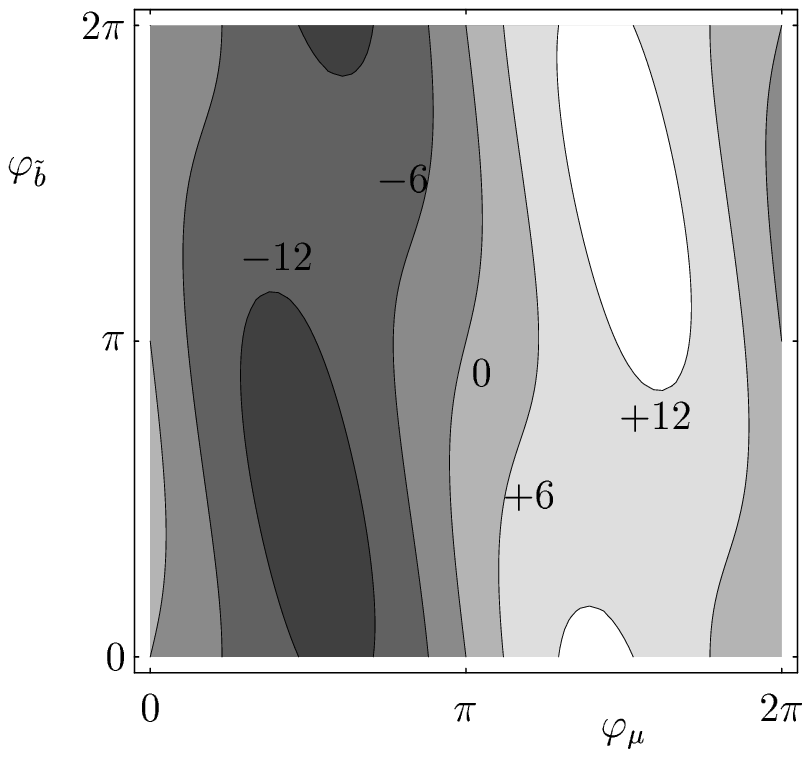}}
\put( 3.35,-0.7){(a)}
\put(11.85,-0.7){(b)}
\end{picture}
\end{center}
\caption{\sl\small The boundaries of the different shaded areas are contour 
lines in the plane $\varphi_{\tilde{f}}-\varphi_\mu$ showing (a) 
Re($d^W_b$[charginos]) in units of $10^{-6}\ \mu_b$ for low 
$\tan\beta$ and (b) Re($d^W_b$[neutralinos]) in units of $10^{-6}\ \mu_b$ for 
high $\tan\beta$. $M_2=|\mu|=m_{\tilde q}=250$ GeV and 
$|m^t_{LR}|=|\mu|\cot\beta$ for (a) and $|m^b_{LR}|=|\mu|\tan\beta$ 
for (b). \label{fig2}}
\end{figure}

\begin{figure}[htb]
\begin{center}
\begin{picture}(15,6.75)
\epsfxsize=8cm
\put(-0.8,-0.2){\epsfbox{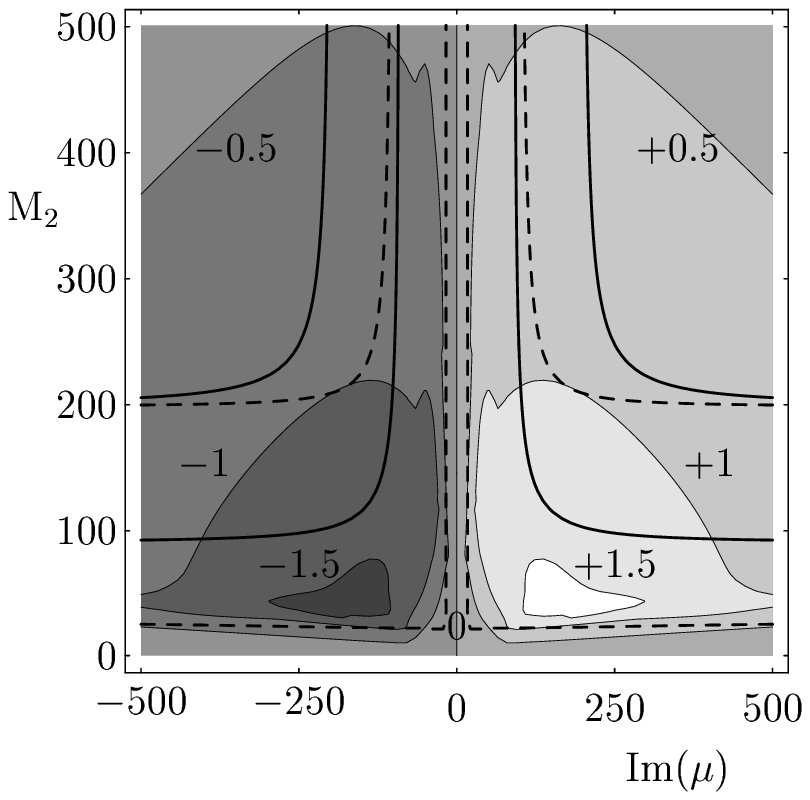}}
\epsfxsize=8cm
\put(7.7,-0.2){\epsfbox{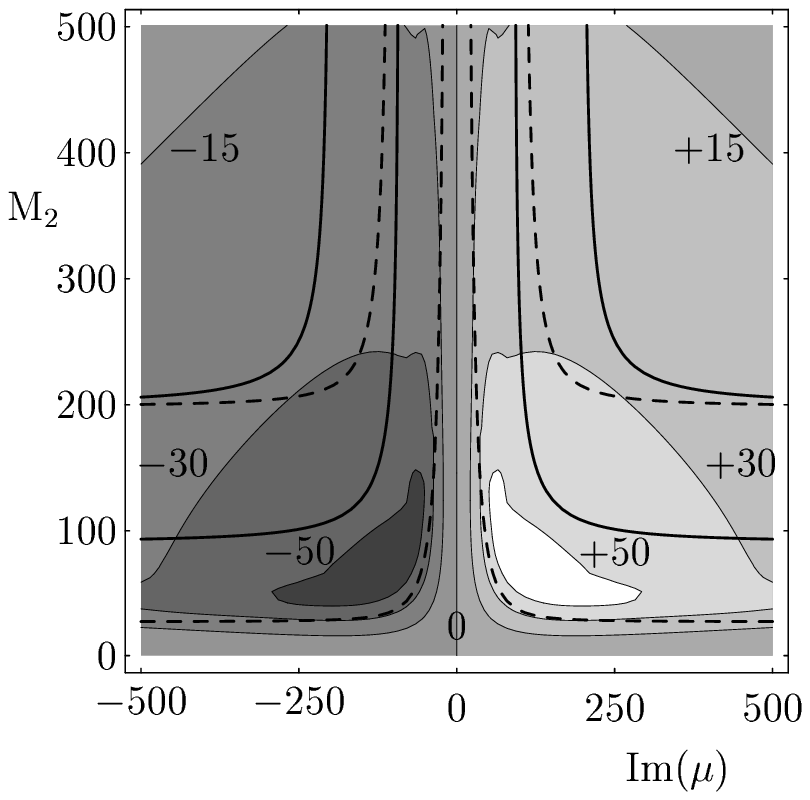}}
\put( 3.35,-0.7){(a)}
\put(11.85,-0.7){(b)}
\end{picture}
\end{center}
\caption{\sl\small The boundaries of the different shaded areas are contour 
lines in the plane $M_2-$Im$(\mu)$ showing Re($d^W_b$[charginos]) in units 
of $10^{-6}\ \mu_b$ for (a) low and (b) high $\tan\beta$ with 
$|\sin\varphi_\mu|=1$, $m_{\tilde q}=250$ GeV and $|m^t_{LR}|=0$. 
Also indicated are the isocurves corresponding to lightest
chargino masses $m_{{\tilde\chi}^\pm_1}=91$ and 200 GeV (solid) and 
lightest neutralino masses $m_{{\tilde\chi}^0_1}=14$ and 100 GeV (dashed). 
The present LEP limits at $\sqrt{s}=183$ GeV are 
$m_{{\tilde\chi}^\pm_1}>91$ GeV and $m_{{\tilde\chi}^0_1}>14$ GeV \cite{janot}. 
\label{fig3}}
\end{figure}

\end{document}